\documentclass[pdflatex,sn-mathphys-num]{sn-jnl}

\usepackage{graphicx}%
\usepackage{multirow}%
\usepackage{amsmath,amssymb,amsfonts}%
\usepackage{amsthm}%
\usepackage{mathrsfs}%
\usepackage[title]{appendix}%
\usepackage{xcolor}%
\usepackage{textcomp}%
\usepackage{manyfoot}%
\usepackage{booktabs}%
\usepackage{algorithm}%
\usepackage{algorithmicx}%
\usepackage{algpseudocode}%
\usepackage{listings}%

\begin{document}

\title[Decoding Musical Evolution Through Network Science]{Decoding Musical Evolution Through Network Science}


\author[1,*]{Niccolò Di Marco}
\author[2]{Edoardo Loru}
\author[3]{Alessandro Galeazzi}
\author[1]{Matteo Cinelli}
\author[1]{Walter Quattrociocchi}

\affil[1]{Department of Computer Science, Sapienza University of Rome}
\affil[2]{Department of Computer, Control and Management Engineering, Sapienza University of Rome}
\affil[3]{University of Padova, Department of Mathematics, Padova, Italy}
\affil[*]{niccolo.dimarco@uniroma1.it}


\abstract{Music has always been central to human culture, reflecting and shaping traditions, emotions, and societal changes. Technological advancements have transformed how music is created and consumed, influencing tastes and the music itself.
In this study, we use Network Science to analyze musical complexity. Drawing on $\approx20,000$ MIDI files across six macro-genres spanning nearly four centuries, we represent each composition as a weighted directed network to study its structural properties.
Our results show that Classical and Jazz compositions have higher complexity and melodic diversity than recently developed genres. However, a temporal analysis reveals a trend toward simplification, with even Classical and Jazz nearing the complexity levels of modern genres.
This study highlights how digital tools and streaming platforms shape musical evolution, fostering new genres while driving homogenization and simplicity.}




\maketitle

\section*{Introduction}\label{sec1}
Music is a defining element of human culture, reflecting and shaping traditions, emotions, and societal changes \cite{bernstein1976unanswered,welch2020impact, cross2001music}. Its capacity to engage cognition and emotion has long intrigued researchers across disciplines \cite{wilkins2014network, zatorre2003music, cross2003music, koelsch2014brain, honing2015without, martinez2016neural,scherer2013music}. Over centuries, it has evolved alongside cultural and technological shifts, adapting to new tools and contexts of creation and consumption \cite{mehr2019universality}.

Historically, music was a communal experience, limited to live performances and closely tied to specific cultural practices \cite{mithen2006singing}. Moreover, composing music was restricted mainly to trained specialists, often working within traditional frameworks.
The invention of new technologies revolutionized this dynamic, allowing music to transcend temporal and spatial boundaries \cite{goodwin2004rationalization}. By the mid-20th century, physical formats like vinyl records and cassette tapes had democratized access to music and enabled a broader audience to participate in music creation, laying the groundwork for today's digital era \cite{brusila2022music,guo2023evolution,frenneaux2023rise,guo2023evolution}. 

This shift has fostered the emergence of new genres and innovative styles, challenging traditional notions of musical expertise and redefining creative boundaries \cite{o2023novelty}.
Moreover, the recent advent of streaming platforms and social networks has reshaped our cultural landscape \cite{DiMarco2024,etta2023characterizing}, including how music is consumed and produced \cite{curien2009music}. 

Platforms like Spotify and YouTube not only offer listeners personalized recommendations but also act as hubs for discovering and promoting new artists, effectively functioning as the ``new radio'' of the digital age \cite{prey2022platform,billboard_spotify_artist,hodgson2021spotify}. Algorithms play a pivotal role in shaping these experiences, tailoring music discovery to individual preferences and thereby influencing listening habits \cite{simon1996designing,  carstens2018social,anderson2020algorithmic,mok2022dynamics}. 

However, this interconnected landscape is not without its drawbacks.
Previous studies have suggested that content circulating in fast, interconnected, and algorithmically curated environments is subject to simplification processes, as seen in the case of song lyrics \cite{parada2024song} and social media comments \cite{dimarco2024vocabulary}. This raises an important question: is a similar trend occurring in the contemporary musical landscape? Addressing this question is challenging due to the lack of a precise method for measuring musical complexity \cite{percino2014instrumentational,febres2017music,chakrabarty2019measuring}.

To fill this gap, we build on the approach of previous studies \cite{kulkarni2024information}, utilizing Network Science tools to analyze musical compositions and examine how democratization and digital connectivity impact musical complexity and diversity.
In particular, we analyze a dataset of approximately $20,000$ MIDI files categorized into six macro-genres \cite{metamidi}, choosing to represent musical compositions as weighted directed networks where notes are nodes and transitions are edges. 
This approach systematically explores structural differences across genres and offers a potential method for measuring musical complexity and its trends over time.

The properties of the networks reveal that Classical and Jazz compositions exhibit greater complexity and melodic diversity than other genres. 
Furthermore, unlike previous studies, our measures offer a detailed characterization of the topological and musical features defining each genre, shedding light on the foundational elements of their identities.
Finally, a temporal analysis of our measures reveals a trend toward simplification in genres such as Classical and Jazz, which have reached complexity levels comparable to those of more recently developed genres, showing lower complexity. 

Overall, our findings suggest that the democratization of music composition and the creation of highly interconnected environments for music sharing have contributed to the emergence of new, less complex, and more homogeneous genres, while even historically more complex genres like Classical and Jazz have experienced a decline in complexity over time.

\section*{Results}\label{sec2}

\subsection*{Genre-Specific Network Properties}
Before the main analysis, we recall that our dataset contains $\approx 20.000$ networks in which notes are nodes and edges are transitions among them.

\begin{figure}[!ht]
    \centering
    \includegraphics[width = 0.8\linewidth]{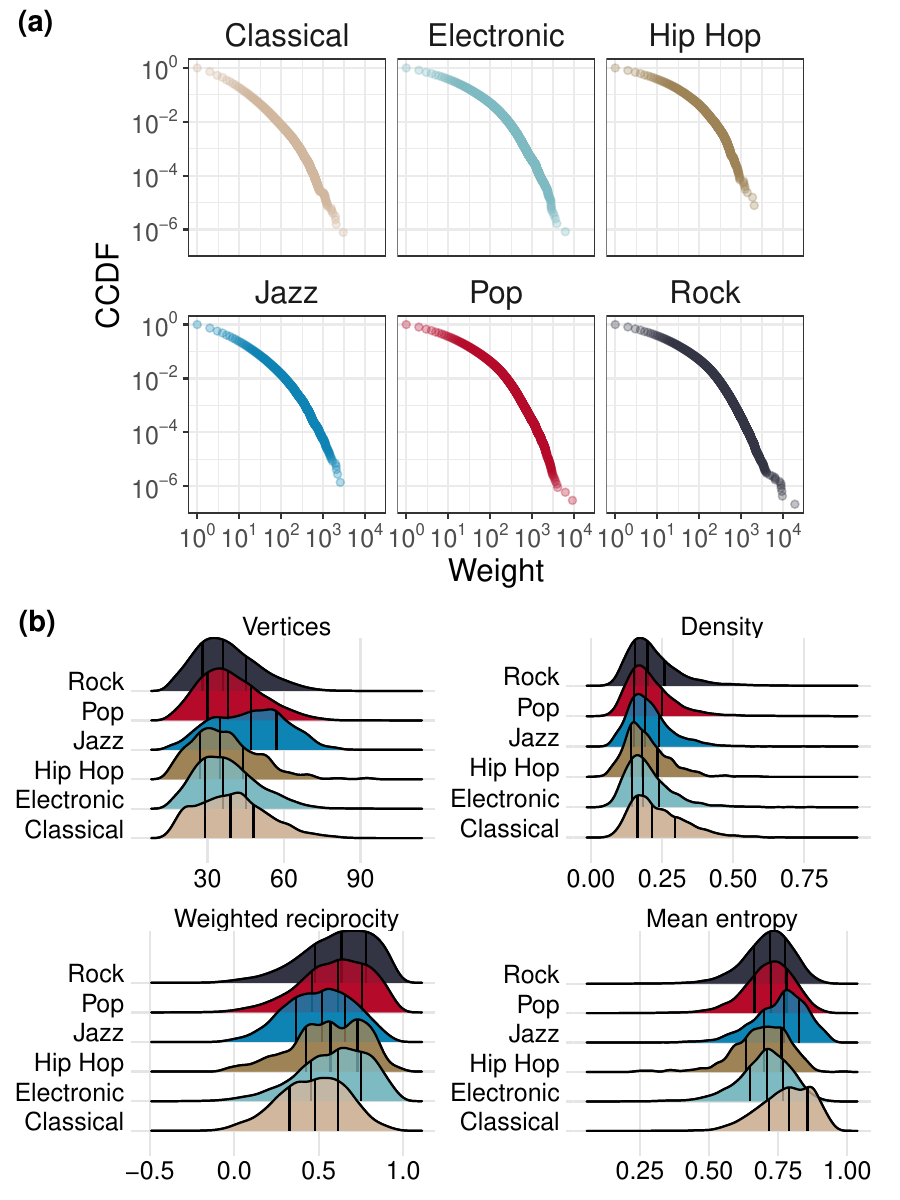}
    \caption{$(a)$ CCDF of aggregate network's weights. $(b)$ distribution of various measures, divided according to the genre. The vertical bars represent the quartiles of each distribution.}
    \label{fig:genre_measures}
\end{figure}

The first natural question is whether networks from different genres exhibit distinct properties. 
To explore this, Fig. \ref{fig:genre_measures} presents several measures calculated on our collection.

Figure \ref{fig:genre_measures}$(a)$ shows the CCDF of all the edge weights appearing in each genre. Notably, all distributions are heavy-tailed, but we observe that Electronic, Pop, and Rock genres exhibit longer tails than Classical and Jazz.

Figure \ref{fig:genre_measures}$(b)$ illustrates the distributions of several network metrics grouped by music genre, namely the number of vertices, density, weighted reciprocity, and mean node entropy. See Methods for each metric's definition and their musical interpretation.

Regarding vertex count, Jazz pieces stand out with a higher number of notes, while other genres consistently exhibit fewer notes. Interestingly, all genres share similar density values. This property may be tied to musical patterns that lead a note to be connected only to a limited number of other notes.

For weighted reciprocity, more pronounced differences emerge: Rock, Pop, Hip Hop, and Electronic music exhibit very high reciprocity, whereas Classical and Jazz show lower median values and, in some cases, even anti-reciprocity. Musically speaking, this suggests that Classical and Jazz may feature extended melodies with non-repeating pairs of notes. Conversely, mainstream genres such as Pop or Rock may favor shorter phrases or `hooks' that can enhance a song's memorability and, in turn, its popularity.

Finally, the higher entropy values observed in Jazz and Classical imply nearly uniform transitions between notes. This indicates that, on average, the transitions between a source node and its targets occur with similar frequency. Interestingly, this points to the fact that musical connections are pre-established among harmonious notes and are likely to occur fairly uniformly.

To test formally the previous insights, we run two-sample Mann–Whitney U tests \cite{mann1947test} between each pair of genres, adjusting the p-value using the standard Bonferroni-Holm correction \cite{holm1979simple}. The results, shown in Fig. S1 in Supplementary Information (SI), are aligned with our insights.

\subsection*{Analyzing Musical Connectivity and Variability}
\begin{figure}[!ht]
    \centering
    \includegraphics[width=0.8\linewidth]{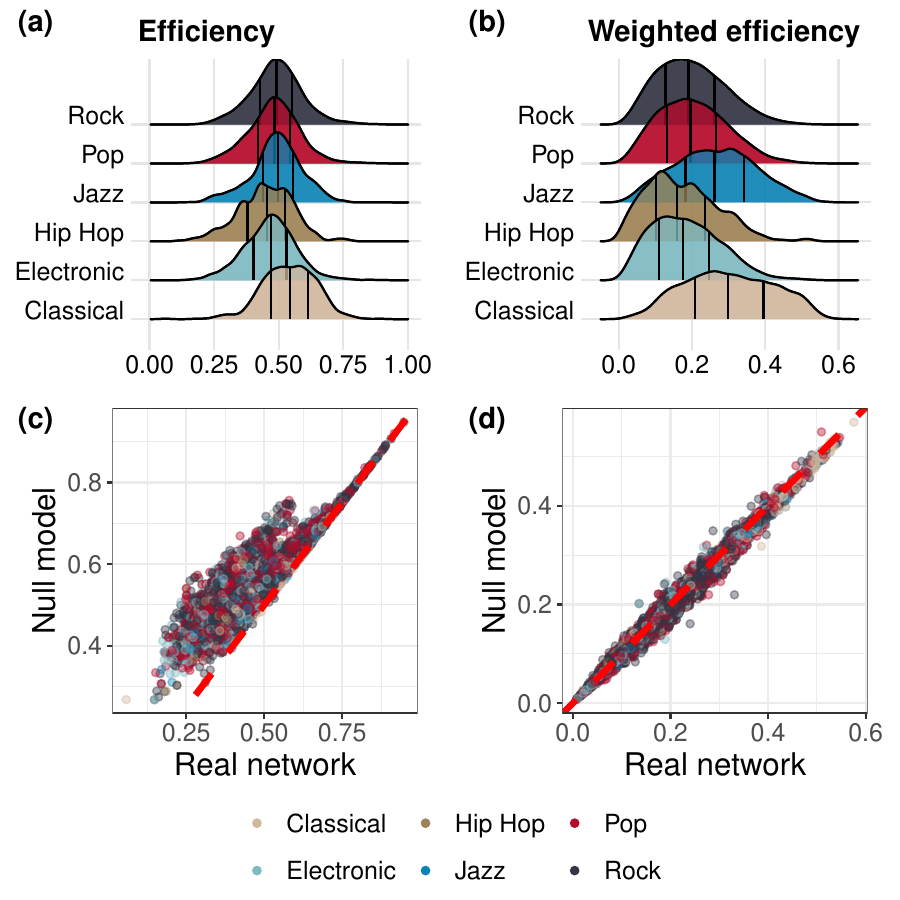}
    \caption{$(a)$ Distribution of unweighted efficiency. $(b)$ Distribution of weighted efficiency.
    $(c),(d)$ comparison between efficiency and weighted efficiency in real and randomized networks. To improve their visualization we use a random sample of $n = 10^4$ points.}
    \label{fig:efficiency}
\end{figure}

In this section, we focus on the spreading properties of networks using the measures of {\it global efficiency} (see Methods section for further details). 
Figure \ref{fig:efficiency} $(a),(b)$ shows the distributions of efficiency and its weighted counterpart for each genre. Recall that these values are connected to the shortest paths of the network and, especially for the weighted case, can be interpreted as measures of musical complexity.

Hip Hop and Classical music stand out among other genres regarding non-weighted efficiency. Specifically, Hip Hop is characterized by lower efficiency values, whereas Classical music typically exhibits higher values. 
These differences become even more evident when weights are considered, i.e. the number of times a transition between two notes occurs. Notably, we observe the greater efficiency of Classical and Jazz structures and lower values of Hip Hop.

Given that all weights satisfy $w_{ij} \geq 1$, higher efficiency corresponds to shortest paths carrying low weights. Since weights represent note transitions, this indicates less repetitive melodies and greater musical variability. Therefore, the high efficiency obtained by Classical and Jazz points to a greater complexity of these genres than the others.

Interestingly, the similarities observed in non-weighted efficiency suggest that while the average distance between two notes is comparable across all genres, their differences lie in the weighted structure of the networks.

To deepen our exploration, we compare the observed efficiency with that derived from ad-hoc null models (see Methods for further details). Briefly, the first null model rewires the edges between notes, preserving the out-degree distribution. The second one shuffles the weights among each node's out-edges, maintaining the out-strength distribution.

The results of this analysis are depicted in Fig. \ref{fig:efficiency} $(c),(d)$.

Panel $(c)$ reveals that, in nearly all cases, the efficiency values in real networks are lower than those obtained from a random topology where edges are rewired but the number of transitions from each note is preserved. Conversely, panel $(d)$ displays values comparable to a null model that preserves the out-strength of each node.

These results highlight some interesting musical patterns. In fact, the edge rewiring procedure allows transitions that may be dissonant, breaking common musical patterns. At the same time, it creates shorter paths, resulting in greater efficiency. Interestingly, this observation suggests that musicality does not necessarily prioritize the formation of short paths.
On the other hand, rewiring only the weights maintains the overall musical structure, changing only the number of transitions between notes. However, the comparable results with a null model may be a consequence of the high values of entropy (shown in Fig. \ref{fig:genre_measures} $(d)$). 

To check the robustness of our results, we repeat the analysis using a different measure, the {\it Network Entropy}, previously used in other work to quantify the information contained in musical networks constructed from J.S. Bach pieces \cite{kulkarni2024information}.
We report the result of the analysis in Supplementary Fig. S2. Notably, we observe similar patterns, with Classical and Jazz music obtaining higher values (i.e. containing higher information) than other genres. For a full discussion on the measure and the results, see Section `Entropy of Networks' in SI.

\subsection* {Embedding Musical Structures in High-Dimensional Spaces}
Up to this point, our analysis has primarily examined musical networks from a purely topological perspective, followed by an interpretation of the results from a musical standpoint. However, this approach may not be sufficient to fully capture the networks' properties. To this end, we also incorporated musical information contained in musical intervals.

Recall that an interval corresponds to the difference in pitch between two notes, measured as the number of semitones between the two. In Western music theory, the main intervals are 12 and range from ``perfect unison'' (0 semitones of difference) to ``major seventh'' (11 semitones of difference). Table \ref{tab:intervals} provides a complete list of these intervals and their corresponding differences in semitones.

\begin{table}[h!]
\centering
\begin{tabular}{l c}
\toprule
\textbf{Interval}       & \textbf{Semitones} \\ \midrule
Perfect Unison                 & 0 \\
Minor Second           & 1 \\
Major Second           & 2 \\
Minor Third            & 3 \\
Major Third            & 4 \\
Perfect Fourth         & 5 \\
Tritone                & 6 \\
Perfect Fifth          & 7 \\
Minor Sixth            & 8 \\
Major Sixth            & 9 \\
Minor Seventh          & 10 \\
Major Seventh          & 11 \\
\bottomrule
\end{tabular}
\caption{Musical intervals and their distances in semitones.}
\label{tab:intervals}
\end{table}

\begin{figure}[!ht]
    \centering
    \includegraphics[width=0.8\linewidth]{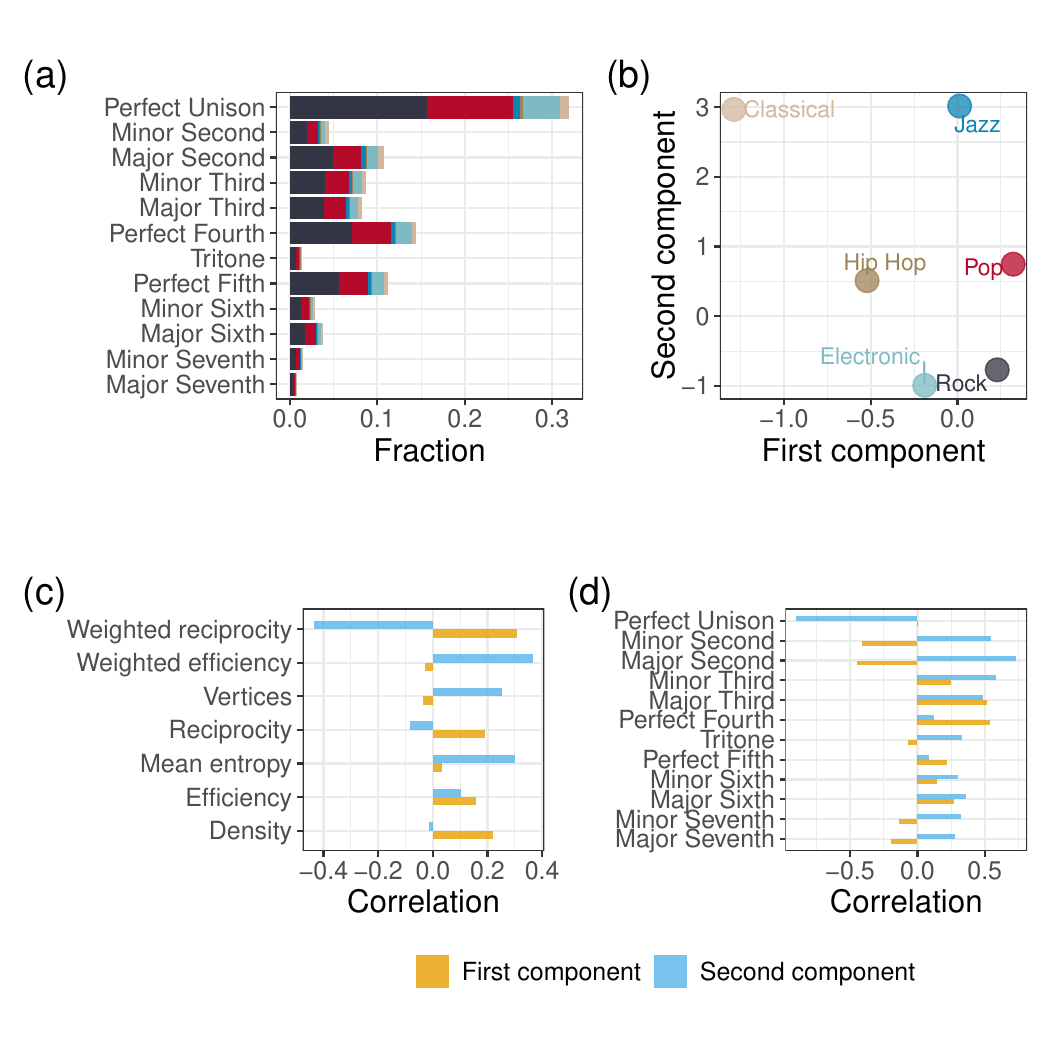}
    \caption{$(a)$ Fraction of intervals appearing in each musical genre. $(b)$ Center of mass for each genre, computed using UMAP $2-$dimensional coordinates. $(c)$ correlation between components and measures. $(d)$ correlation between intervals entries and components.}
    \label{fig:interval_embeddings}
\end{figure}

Note that these intervals ignore the octave the two notes are in. How frequently these intervals occur within a song reflects its musical properties, such as its key and the mood it may evoke in a listener. For instance, in music that might be generally described as `sad', so-called `minor' intervals will be predominant. Conversely, `happy' music is typically characterized by extensive use of `major' intervals. To give an idea of intervals appearing in our dataset, Fig. \ref{fig:interval_embeddings}$(a)$ shows the fraction of intervals appearing in songs, divided by genres. Notably, our results are coherent with those obtained in previous work \cite{mehr2019universality}.

In the section `Network Embeddings' of methods, we explain in detail how we construct interval embeddings of our dataset.
Briefly, we associate to each $G$ a vector $v_G$ whose components are linked to the number of times a specific interval appears in the song, not distinguishing among notes at different octaves.

To check the robustness of these results, in SI we repeat the analyses of this section using \texttt{graph2vec}, an algorithm that embeds networks in a high dimensional space using their topological properties. Notably, we observe consistent outcomes, demonstrating that our findings are robust and can be achieved using either topological or musical properties. We also compare the embeddings of real networks with the ones created from randomized versions of networks, observing clear differences between the two. These results corroborate the efficacy of our procedure in capturing relevant musical properties of networks. Finally, in the Section `Analysis with popularity' of SI, we add analyses involving the popularity of songs and artists, based on network embeddings.

Figure \ref{fig:interval_embeddings} $(b)$ shows the center of mass of each genre, computed starting from the $2-$dimensional representation of UMAP \cite{Konopka2018}. 
We report the plot containing all the points in Supplementary Fig. S4. 

Although most genres appear mixed up, a clear cluster for Classical music emerges, highlighting its difference from other genres. These differences become even clearer in panel $(b)$, revealing a notable distance between Jazz and Classical from other genres.

To understand what network properties each component embeds, in panel $(c)$ we show the Pearson correlations between the coordinates of songs on the UMAP plane and their measured network properties. 
The first component shows significant correlations only with topological properties such as efficiency ($r \approx 0.16, p < 0.001$), density ($r \approx 0.22, p < 0.001$), and reciprocity ($r \approx 0.18, p < 0.001$). On the other hand, the second component is correlated mostly with weighted properties such as weighted efficiency ($r \approx 0.37, p < 0.001$), mean entropy ($r \approx 0.30, p < 0.001$), and weighted reciprocity ($r \approx -0.44, p < 0.001$). 
While the first component appears to reflect a combination of topological properties, the second one captures the weighted structure and complexity of the song. These findings align with the results presented in Fig. \ref{fig:genre_measures}.
Moreover, while Classical and Jazz share similar values on the second component (i.e. similar complexity), they differ on the first, suggesting that their differences may be rooted in topological properties. 

Conversely, to assess which intervals contribute to the two components, we compute the correlations between the coordinates and each component of the $12-$dimensional vector. 
Mathematically speaking, let us consider the $N \times 12$ matrix $M = [{\bf v_j}]_{j = 1\ldots 12}$, where $N$ is the number of songs and each row represents the vector associated with a composition. We compute the correlations $r({\bf v}_j, pr_i), j = 1 \ldots 12, i = 1,2$ where $pr_i$, denotes the coordinates of $i-$th component according to UMAP. 

The resulting values are reported in Fig. \ref{fig:interval_embeddings}$(d)$.

We find that the first component is strongly influenced by the presence of major thirds and perfect fourths, while the second component is predominantly associated with minor and major seconds, as well as unisons. 
These relationships help identify which intervals are more characteristic of specific genres.

For instance, both Classical and Jazz compositions frequently rely on unisons and minor/major seconds and thirds, but they differ in their emphasis on major thirds and perfect fourths. Similarly, genres like Hip Hop-Pop, Electronic, and Rock exhibit comparable patterns in their interval distributions, though with distinct nuances.

A further interpretation of the two dimensions comes from Western music theory, where intervals are commonly classified as either `stable' or `unstable' based on their tendency to resolve. Resolution refers to moving from a state of dissonance that might evoke incompleteness or suspense, to one of consonance that feels definitive and accomplished. Stable intervals, such as the unison and perfect fifth, are consonant and do not require resolution, providing a sense of repose. Imperfect consonances like major and minor thirds and sixths are also stable but slightly less complete. In contrast, unstable intervals, including major and minor seconds, major and minor sevenths, and the tritone, create tension and seek resolution to more stable intervals, such as thirds or octaves. With these notions in mind, we observe that the first component is negatively correlated with unstable intervals and positively correlated with stable ones. Conversely, the second component exhibits mostly stronger correlations with unstable intervals and weaker correlations with stable ones.

\subsection* {Tracing Musical Evolution Over Time}
In the previous sections, we analyzed the networks' properties without considering the release period of each song. Here, we incorporate this additional dimension to explore the temporal evolution of musical pieces.

Since the Spotify API often provides incorrect release dates —due to associations with remastered or reissued albums— we developed a heuristic approach using the LLM Gemini to approximate the release date of each song (details in the 'Release Date Collection' section of Methods).
Our method assigned release dates to $15,192$ songs, covering approximately $72\%$ of the dataset.

\begin{figure}[!ht]
    \centering
    \includegraphics[width=0.8\linewidth]{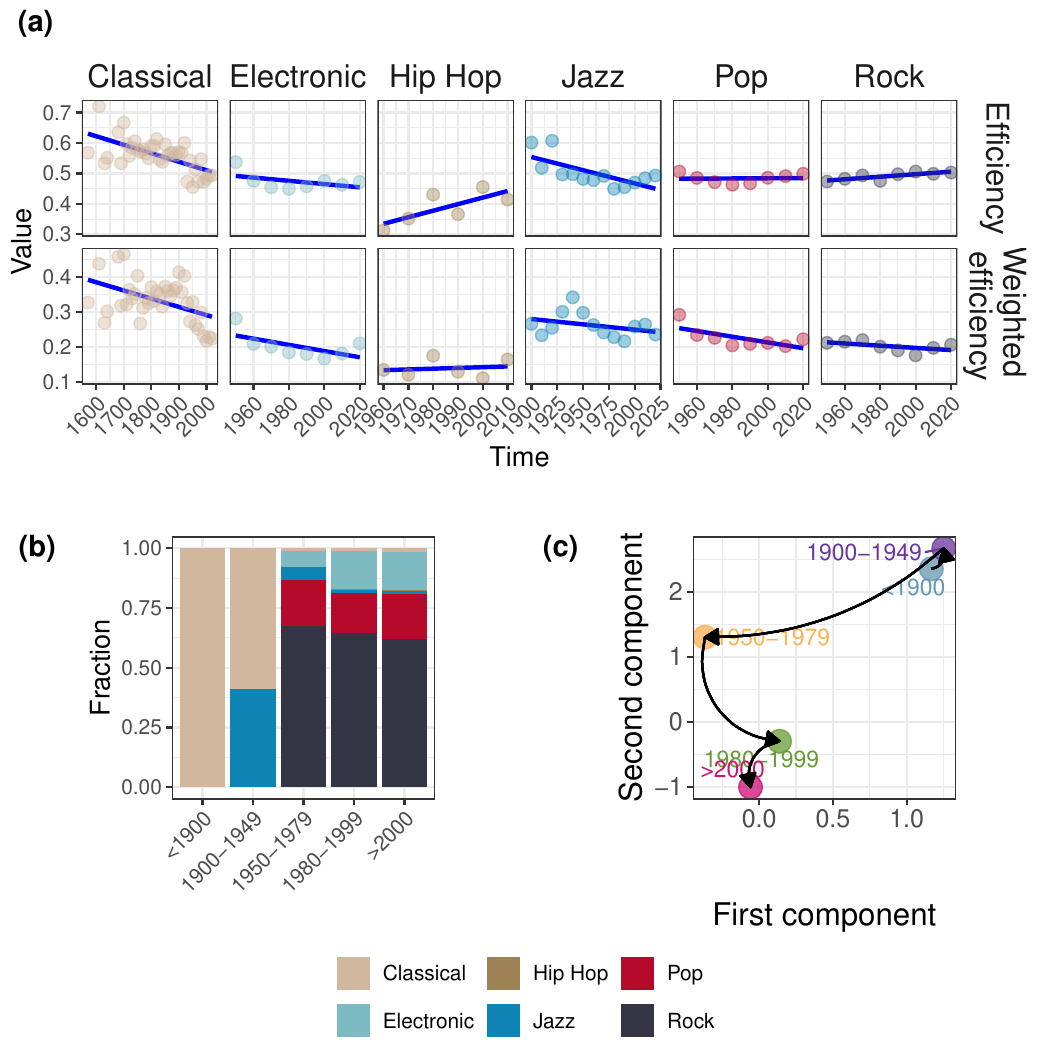}
    \caption{$(a)$ Evolution of mean efficiency measures over decades. The arrows highlight the temporal evolution of considered eras. $(b)$ Distribution of genres in each musical era. $(c)$ Center of mass of each musical period, obtained using UMAP on the interval embeddings.}
    \label{fig:time_evolution}
\end{figure}

Key findings are presented in Fig. \ref{fig:time_evolution}.

In particular, panel $(a)$ displays the decade-averaged efficiency values for each genre. Notably, Classical music exhibits a declining trend, whereas Jazz shows an initial increase in its complexity in its early days, followed by a decline and eventual stabilization. In contrast, the other genres maintain relatively flat patterns, with efficiency values comparable to those of Classical and Jazz in recent years.
To formally validate these observations, Table S1 in SI presents the results of Mann-Kendall tests applied to each trend.

This result suggests a decrease in the complexity of Classical and Jazz music, while the other genres maintain (on average) the same complexity, resulting in values comparable with the ones obtained by Jazz and Classical in recent years.

To focus on the musical properties of the MIDIs, we employ the interval embeddings introduced in the previous section. Moreover, we associate each song to one between $5$ musical periods, namely $<1900,1900-1949,1950-1979,1980-1999,>2000$. 
Figure \ref{fig:time_evolution}$(b)$ shows the genre distribution among those classes. In particular, we observe a transition from Classical and Jazz music to a prevalence of popular genres such as Rock, Pop and Electronic.

Finally, panel $(c)$ shows the $2-$dimensional coordinates center of mass of each musical era, where coordinates were obtained using UMAP on the embeddings. The arrows indicate the temporal evolution between musical periods, highlighting a trend starting from the up-right part of the plot through the down-left quadrant.

To interpret the meaning of the two components, Figure S9 in SI provides the correlation between components with measures and intervals occurrence, respectively.
The components maintain a similar interpretation of Figure \ref{fig:interval_embeddings}. The first component anticorrelates with topological properties, while the second one is correlated with complexity and weighted properties.

Again, the results highlight a greater complexity (and similarity) for music composed before $1950$, and the same holds for recent music, even with lower complexity. Notably, we can observe a transition that, starting from higher complexity, tends to simplicity. At the same time, the differences in the first components may suggest a transition also to different topological properties. Finally, it is interesting to note how recent music shares similar properties, with music composed in the $1950-1979$ period acting as a bridge between the two eras.

The observed results could partly be attributed to the rise of more homogeneous and less complex genres in recent years. However, when considered alongside the findings presented in panel $(a)$, our analysis indicates that even enduring genres like Classical and Jazz have undergone a noticeable simplification compared to their origins.

Overall, our study highlights that the democratization of the composition process and the advent of new technologies and platforms have fostered the development of genres characterized by reduced complexity relative to earlier eras.

\section*{Discussion}

This study applies the tools of Network Science to provide a novel perspective on the analysis of music, offering insights into how structural properties of musical compositions vary across genres, correlate with popularity, and change over time. By representing musical compositions as networks, we provide a methodological framework that can be extended to other cultural domains. Moreover, our findings open avenues for interdisciplinary research, bridging musicology, data science, and sociology to investigate how digital environments shape creativity and consumption.

Despite the novel insights, our study is subject to several limitations. First, reliance on the MetaMidi Dataset introduces potential inaccuracies in genre classification. Moreover, while genre tagging is helpful for broad categorizations, it may oversimplify the diversity within and across genres, particularly in hybrid or experimental music.

Second, linking MIDI files to Spotify entries required heuristic methods. Although these heuristics were carefully designed, mismatches or omissions could result in occasional inaccuracies in metadata associations. This issue is especially pertinent for older or less popular tracks, where metadata availability tends to be sparse or inconsistent.
Third, estimating release dates posed a significant challenge since Spotify metadata often fails to reliably indicate the original release dates. To address this, we employed a large language model (LLM) to infer release dates. While innovative, this approach introduces an additional layer of uncertainty, as the LLM relies on contextual data that may be occasionally inaccurate or comprehensive.
Lastly, our focus on MIDI data inherently limits the scope of our analysis to structural aspects of music, such as note transitions and melodic complexity. Other critical dimensions, such as lyrics, timbre, production techniques, and cultural context, remain unexplored. Future research could integrate these elements to provide a more holistic understanding of musical evolution. 

Despite these limitations, our findings offer several important implications. In fact, the observed trend of musical simplification reflects broader societal changes, including the influence of global interconnectedness, rapid content dissemination, and the algorithmic curation of music consumption. 

By situating these findings within the broader context of technological and societal change, we provide a foundation for future research exploring the interplay between creativity, culture, and technology. 

\backmatter

\section*{Methods}

\subsection*{Dataset}
\subsubsection*{MIDI collection}

The Musical Instrumental Digital Interface, in short MIDI, is a widely used standard for representing musical information in a digital format. Rather than storing audio, the MIDI format encodes musical events such as note pitches, durations, and timing, along with performance controls such as tempo changes. Specifically, MIDI data is organized into channels, allowing multi-instrument encoding. This representation renders it particularly suitable for music information retrieval tasks.

In this study, we employ the MetaMIDI Dataset (MMD) \cite{metamidi}, a publicly accessible collection of more than 400,000 MIDI files. Specifically, we focus on the subset containing MIDI files that are annotated with a title, artist name, and at least one music genre. 
We point to the dataset's official GitHub repository \cite{metamidi_github} for further details about its construction. 

From this initial sample of approximately $160,000$ MIDI files, we keep only tracks with genres containing at least one keyword between ``rock'', ``pop'', ``electronic'', ``classical'', ``jazz'', ``hiphop'', or ``hip hop''. Note that a single track from this list may be associated with multiple genres. We further refine this selection by keeping only MIDIs longer than 60 seconds that can be successfully parsed by the R library \texttt{tuneR} \cite{tuneR}. This filtering procedure results in approximately $40,000$ MIDI files.

\subsubsection*{Spotify metadata}
To enrich our dataset, we conduct a data collection phase on Spotify using the R package \texttt{spotifyr} \cite{spotifyr}, a wrapper for the official Spotify API.
First, we preprocess the title and artist associated with each MIDI removing:

\begin{enumerate}
    \item everything that appears after the word feat. of ft.;
    \item everything appearing between parentheses;
    \item all the non-alphanumerical characters.
\end{enumerate}

Then, for songs with multiple authors ($\approx 22.15 \%$ of our dataset), we select only the first one. The rationale behind this choice is that, in most classical pieces, the original composer is listed first, followed by the names of performers. Moreover, in the main analysis, we usually do not consider artists, reducing the impact of this choice.

Using these cleaned songs and artist names, we gather information from Spotify. In particular, for each artist in our dataset, we use the \texttt{search\_spotify} function to retrieve their Spotify ID, name, popularity, number of followers, and associated genres. This process is repeated for every unique artist.

Next, for each song $s$ composed by artist $a$, we use the \texttt{search\_spotify} function using the query ``track: $s$, artist: $a$'', that returns a set of maximum $50$ songs matching the query. To identify the correct one, we first filter for tracks where $a$ is listed as one of the composers. Then, we select the song whose name has the smallest Levenshtein distance from $s$. If multiple songs meet this criterion, we select the one with the earliest album release date to avoid remastered or live versions. By doing so, the MIDI files are matched with their corresponding Spotify metadata, including the ID, name, popularity, and the release date of the album they belong to.

To ensure that only distinct songs are considered, we select a subset of the MIDI files with unique Spotify IDs, resulting in a final dataset of 21,480 unique songs. Extended Data Figure \ref{fig:dataset} shows the distribution of genres in this final dataset.

\subsection*{Network construction}
Starting from our Dataset, we use {\it tuneR} library to read MIDI files as data tables containing (among others) time, note, channel, and track of each MIDI event. This allows us to construct directed networks in which nodes are distinct notes and edges indicate a transition between the two nodes, weighted by how many times it occurs. More in detail, suppose that $x,y$ are two notes played in a certain MIDI file. A directed edge between $x$ and $y$ means that $y$ follows $x$ at least once, with the weight $w_{xy}$ counting how many times this specific transition occurs.

In the main analysis, to focus only on transitions between different notes, we delete all loops, i.e. we set $w_{xx} = 0$ for all $x \in V$, thus obtaining simple networks. In the case of chords, i.e. multiple notes played together, edges are drawn between all the notes in the first and second chord, as in a complete bipartite graph. Note that we apply this procedure to each channel separately, to avoid mixing up different instruments. Further, we ignore the channel associated with drums. 

\begin{figure}[!ht]
    \centering
    \includegraphics[width = 0.7\linewidth]{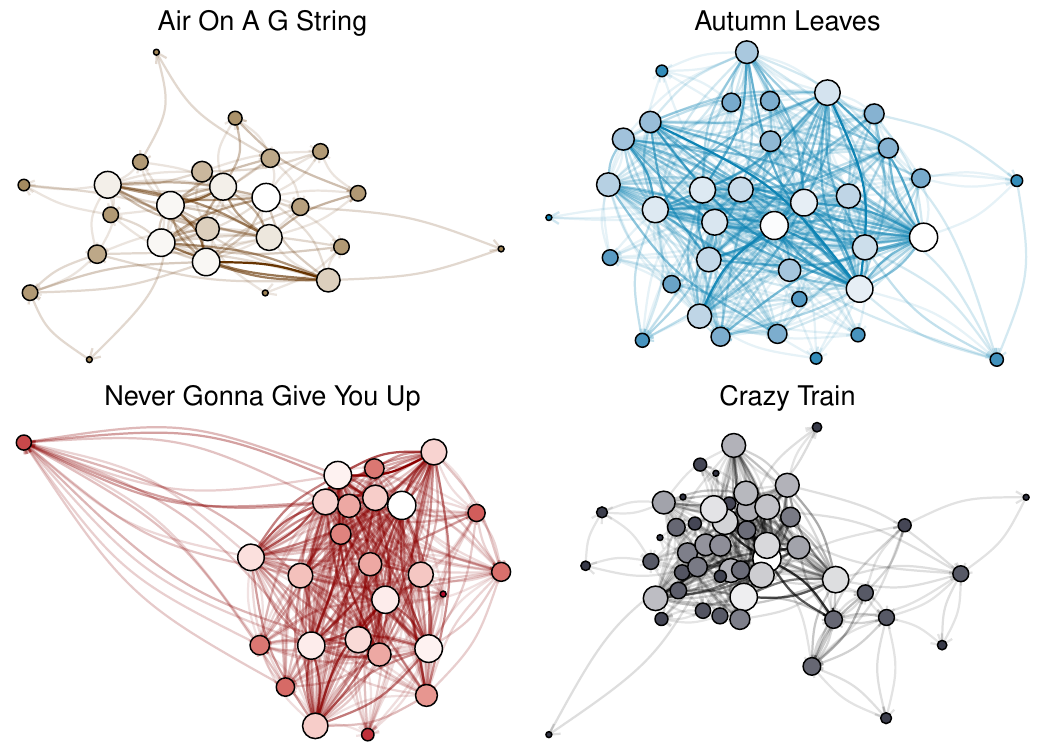}
    \caption{Networks constructed from four MIDI files. The size of each node is proportional to the degree of the node and the transparency of each edge is proportional to its weight.}
    \label{fig:network_examples}
\end{figure}

\begin{figure}[!ht]
    \centering
    \includegraphics[scale = 0.5]{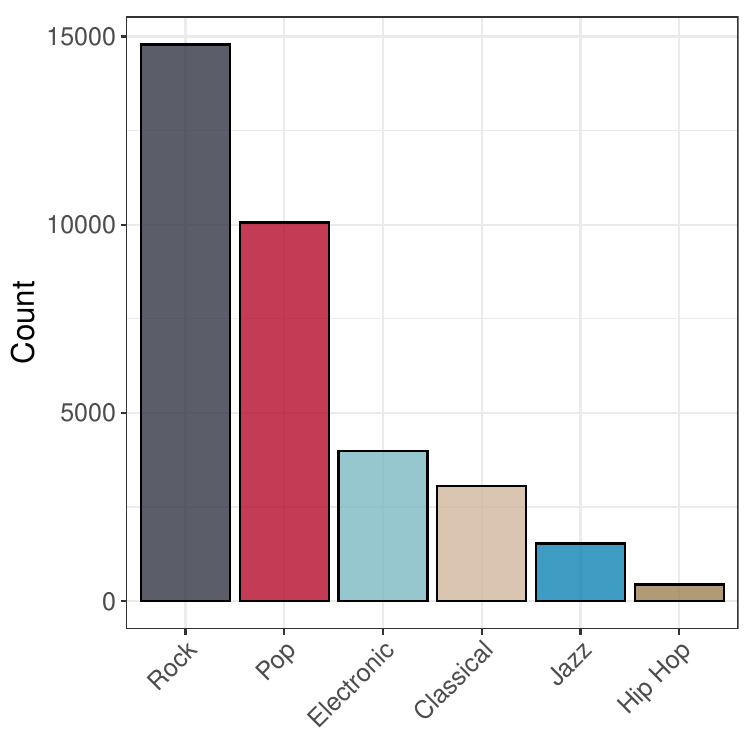}
    \caption{Distribution of genres in our dataset. Recall that one MIDI may be associated with multiple genres.}
    \label{fig:dataset}
\end{figure}

Figure \ref{fig:network_examples} shows examples of networks gathered from different musical genres.

\subsection*{Null models}\label{sec:null_model}
We adopt two different models, one preserving the degree distribution and the other keeping the strength distribution of nodes. 

In the first case, we apply a rewiring of the edges, without considering the weights, akin to the procedure described in \cite{maslov2002specificity}. In this way, we maintain the number of transitions that start between each note, but we change how they are connected, possibly breaking common musical patterns. 

In the latter case, we use a model previously used in many works involving weighted directed graphs \cite{Opsahl_rich_club,alipour2024drivers}. This model locally reshuffles the weights of the outgoing links, without changing the connections. 
Note that, from a musical point of view, this model keeps all musical choices (such as melodies or chords) intact, changing only the number of transitions between note pairs. 

\subsection*{Networks' embeddings}
To obtain a comprehensive picture of the properties of the MIDIs, we embed each network into a high-dimensional space using two different approaches. The first approach associates each network $G$ to a $12-$dimensional vector $v_G$ in which each position is associated with a musical interval (see Table \ref{tab:intervals}). In particular, $v_G[i]$ contains the number of times the interval $i$ appears in the MIDI. For comparison purposes, we normalize each vector by requiring $|| v_G ||_2 = 1$. In this space, two networks are close if they share similar intervals, suggesting similarity from a musical point of view.

The second approach involves using \texttt{graph2vec} \cite{narayanan2017graph2vec}, a well-established algorithm to create network embeddings. Specifically, we employ the implementation available in the Python package \texttt{karateclub}. The results using this latter approach are in SI.

We highlight that these two methods capture completely different network properties. Indeed, the former relies on purely musical features, whereas the latter exploits the topological properties of the networks. 

\subsection*{Network measures}
Networks are widely used across various fields and numerous metrics have been developed to measure their properties. In this section, we recall some measures, focusing on their interpretation in our context.

\vspace{1mm}

\emph{Density:} the density of a network is defined as its number of edges over the maximum possible i.e. $d = \frac{|E|}{|V||V-1|}$. In particular, the density measures how many note transitions have occurred compared to the possible ones. Hence, high values may suggest higher complexity, even if the measure itself does not contain any information about the type of transition.

\vspace{1mm}

\emph{Reciprocity:} in a directed network, an edge $(i,j) \in E$ is {\it reciprocated} if also $(j,i) \in E$. However, instead of simply computing the fraction or reciprocated edges $r$, it is common \cite{garlaschelli2004patterns} to define reciprocity as   

\begin{equation}\label{eq:reciprocity}
    \rho = \frac{r - a}{1 - a}   
\end{equation}

where $a$ is the network density. This measure is bounded in $[-1,1]$, where positive values suggest that reciprocated links appear more than expected at random. The opposite is true when $\rho < 0$. Values close to $0$ suggest a behavior comparable to a random model.

To take into account the role of edges' weight we adopt the notation and measures introduced in \cite{squartini2013reciprocity}.
In particular, the authors define the {\it weighted reciprocity} as

$$
r = \frac{W^\leftrightarrow}{W}
$$

where $W$ is the sum of the weights of the network, while $W^\leftrightarrow$ is the total reciprocated weight. Without delving further into the definitions, the interested reader can find more information in \cite{squartini2013reciprocity}.

Based on the previous measures, we compare the value of $r$ with those of a null model computing 

$$
\rho_w = \frac{r - r_{NM}}{1 - r_{NM}}
$$

where $r_{NM}$ is the value obtained in a randomized network following the procedure highlighted in Section `Null Models'. Note that the interpretation of the measure's value is the same as \eqref{eq:reciprocity}.

\vspace{1mm}

\emph{Mean node entropy:} entropy is a widely used measure of distribution concentration, originally defined by Shannon \cite{shannon2001mathematical}. For a random variable \( X \) with image \( \mathcal{X} \), entropy is computed as:

$$
H(X) = - \sum_{x \in \mathcal{X}} p(x) \log p(x),
$$

where \( p(x) = P(X = x) \).
In the case of a distribution concentrated at a single point (i.e., a delta-like distribution), entropy reaches its minimum value, \( H(X) = 0 \). Conversely, a uniform distribution achieves the maximum possible entropy, \( H(X) = \log(n) \), where \( n = | \mathcal{X} | \). For comparative purposes, we normalize \( H(X) \) to the interval \( [0,1] \) by dividing it by its maximum value, \( \log(n) \).
We employ entropy to measure the degree of heterogeneity in the out-weight distribution of each network node.

In particular, let us denote with $X_i$ a random variable having image the out-weights distribution of node $i$. For a given network $G$, we compute $\Bar{H} = \frac{1}{N} \sum_{i = 1}^N H(X_i)$ to measure the expected node heterogeneity.

Values close to $1$ suggest that transitions among connected notes occur almost uniformly, while lower values indicate a preference through a subset of the connected notes (i.e. higher probabilities of transition). 

\vspace{1mm}

\emph{Global Efficiency:} the {\it global efficiency} \cite{latora2001efficient} of a network $G$ is a measure of how well the network can efficiently spread information in parallel, and it is defined as

\begin{equation}
    E(G) = \frac{1}{|V||V-1|} \sum_{ij} \frac{1}{d_{ij}}
\end{equation}

where $d_{ij}$ is the length of the shortest path between $i$ and $j$. Note that $E \in [0,1]$, where $E \approx 1$ indicates very well-connected (i.e. low distances) nodes.

For weighted networks, $d_{ij}$ is the shortest weighted path between $i,j$. Since our networks have integer weights, the measure remains bounded in $[0,1]$, maintaining the same interpretation.

In the case of musical networks, high weighted efficiency values are obtained when most of the occurrence between nodes appears a limited number of times (i.e. shorter weights paths), suggesting pieces with non-repetitive melodies and higher musical variance. 

\subsection*{Release date collection}
In some cases, Spotify may report release dates that differ from the actual ones. This is especially common for remastered or reissued albums, where the reported release date corresponds to the newer version rather than the original release. Such differences may become an issue when analyzing the complexity of musical pieces or their popularity over time. 

To address it, we developed a heuristic based on prompting a Large Language Model (LLM), specifically Gemini, developed by Google. In detail, we perform queries to model ``gemini-1.5-flash'' via the official API, which is freely accessible (\href{https://ai.google.dev/gemini-api/}{https://ai.google.dev/gemini-api/}), and ask the model to respond with the release date of a song given its name and its artist's name. Additionally, we ask the model not to provide a date for songs it does not know about.

Since LLM outputs involve an inherent element of randomness and are known to contain so-called ``hallucinations'', to minimize the impact of the model errors we set a threshold on the release date of each genre. Specifically, for Rock, Pop, Electronic, and Hip Hop, we included only songs released after 1950. For Jazz, we included songs released after 1900. Moreover, we also delete all songs incorrectly classified after $2021$, since the dataset was released in that year.

Finally, to check the robustness of our procedure we validate the release dates provided by Gemini and Spotify against a manually annotated sample of $100$ songs. 
The results, presented in SI, demonstrate that Gemini surpasses Spotify's release date estimation only for older songs, specifically those released before 1980.

For this reason, we have chosen to use Gemini's estimates for songs released before 1980, while retaining Spotify's release date for more recent songs, thereby taking advantage of the strengths of both methods.

\bmhead{Acknowledgements}
The work is supported by IRIS Infodemic Coalition (UK government, grant no. SCH-00001-3391), 
SERICS (PE00000014) under the NRRP MUR program funded by the European Union - NextGenerationEU, project CRESP from the Italian Ministry of Health under the program CCM 2022, PON project “Ricerca e Innovazione” 2014-2020, and PRIN Project MUSMA for Italian Ministry of University and Research (MUR) through the PRIN 2022. 
A.G. thanks CY4GATE for the financial support.

\subsection*{Data availability}\label{subsec:datadecl}
This work uses the MetaMidi dataset, a dataset accessible upon request \cite{metamidi}.

\subsection*{Author contributions}
Conception: N.DM.; Experiment design: N.DM., E.L.; Data collection: N.DM., E.L.; Data analysis: N.DM., E.L.; Draft manuscript preparation: N.DM., E.L., A.G.; Code writing: N.DM., E.L.;
All the authors provided useful feedback and revised the manuscript.

\subsection*{Competing interest declaration}
The authors declare no competing interests.

\subsection*{Additional information} 
\textbf{Correspondence and requests for materials} should be addressed to Niccolò Di Marco. \\

\section*{Extended data}
\renewcommand{\figurename}{Extended Data Figure}
\renewcommand{\tablename}{Extended Data Table}

\clearpage
\section*{Supplementary information}

\subsection*{Statistical tests}
In this section, we describe the procedure applied to compare different distributions and discuss their results. In particular, to assess if there are differences in the distributions of Fig. 1$(b)$, we employ a two-sample Mann–Whitney U test \cite{mann1947test}. To account for multiple comparisons, we employ the standard Bonferroni-Holm correction \cite{holm1979simple}.

\begin{figure}[!ht]
    \centering
    \includegraphics[width=0.8\linewidth]{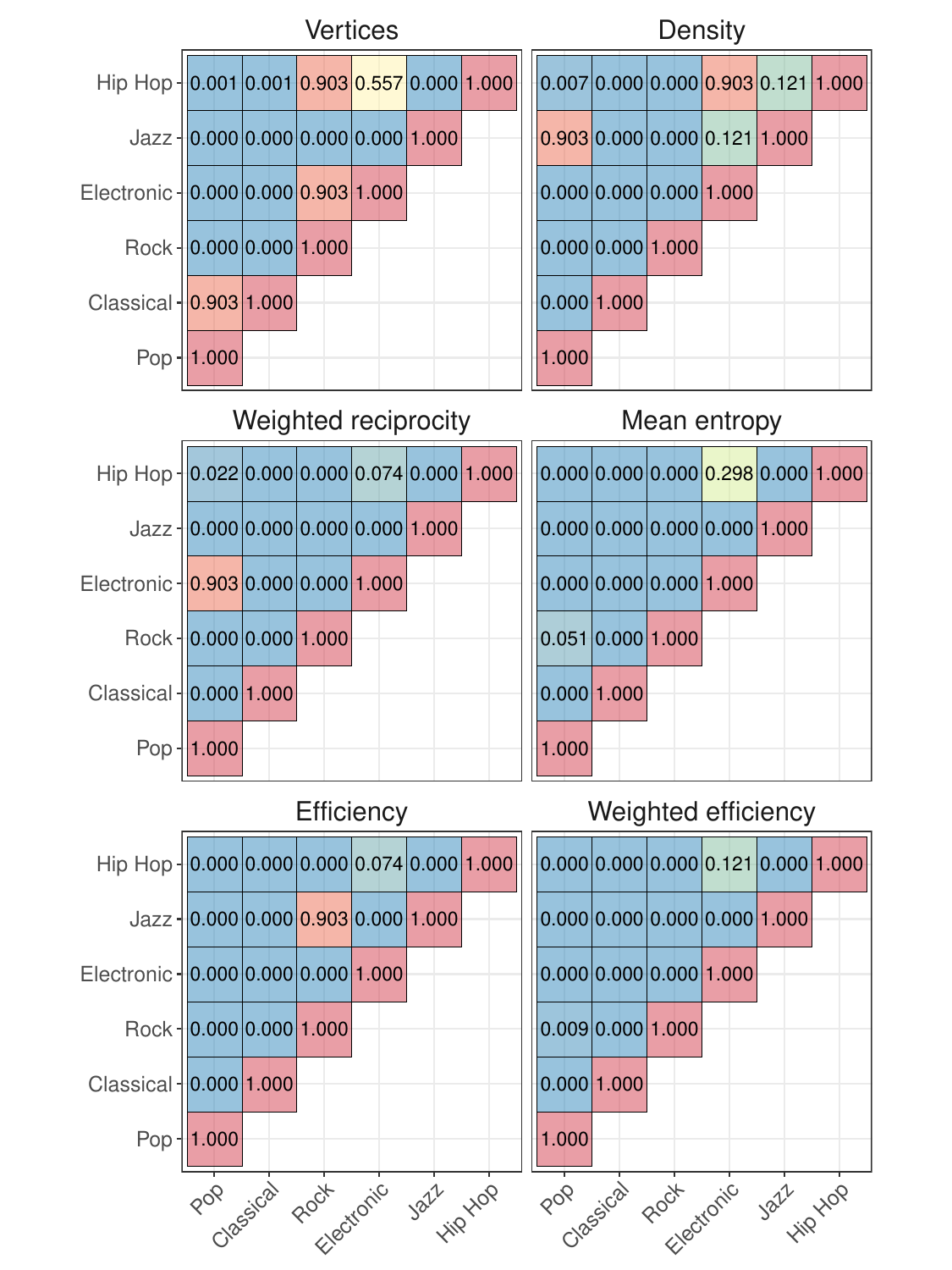}
    \caption{Corrected $p-$values of the Mann-Whitney U test for distribution coming from different genres.}
    \label{fig:basic_measures_tests}
\end{figure}

The corrected $p-$values resulting from coupled tests are depicted in Fig. \ref{fig:basic_measures_tests}.

Even if most tests reject the null hypothesis, there are several similarities between Rock, Pop, Electronic and Hip Hop genres. Classical music exhibits a distribution comparable with other genres only for vertex count, while Jazz music shows comparable values of Density and Efficiency with Pop and Rock music, respectively.

\subsection*{Entropy of Networks}
To study the spreading properties of a network and its overall information, it is common to focus on the Markov chain associated with its structure.
In more detail, we can measure the information contained in node transitions using the Shannon Entropy \cite{shannon1948mathematical}. 
Formally, let us consider a weighted directed network $G$ and denote as $P$ its associated stochastic matrix. The entropy at the node level is:

\begin{equation}
    H_i = - \sum_{j} P_{ij} \log P_{ij}.
\end{equation}

Instead, to compute the entropy of the whole network, it is necessary to weight the contribution of each node $i$ by the stationary distribution of each node $\pi_i$ \cite{meyn2012markov}, that is:

$$
H = \sum_i \pi_i H_i.
$$

Unfortunately, for directed networks, there is no close form for the stationary distribution $\pi$, which instead depends on the specific structure of the network \cite{serfozo2009basics} and may not be unique in the case of non-strong connected networks.
To guarantee the existence and uniqueness of $\pi$ in such cases, we add a small damping probability akin to the page rank procedure \cite{brin1998anatomy}, i.e. we correct $P$ considering

$$
\bar{P}_{ij} = (1-\alpha) P_{ij} + \alpha \frac{1}{n},
$$

using $\alpha = 0.05$. We then compute numerically $\pi$ and the entropy of our collection of networks.

\begin{figure}[!ht]
    \centering
    \includegraphics[width=0.8\linewidth]{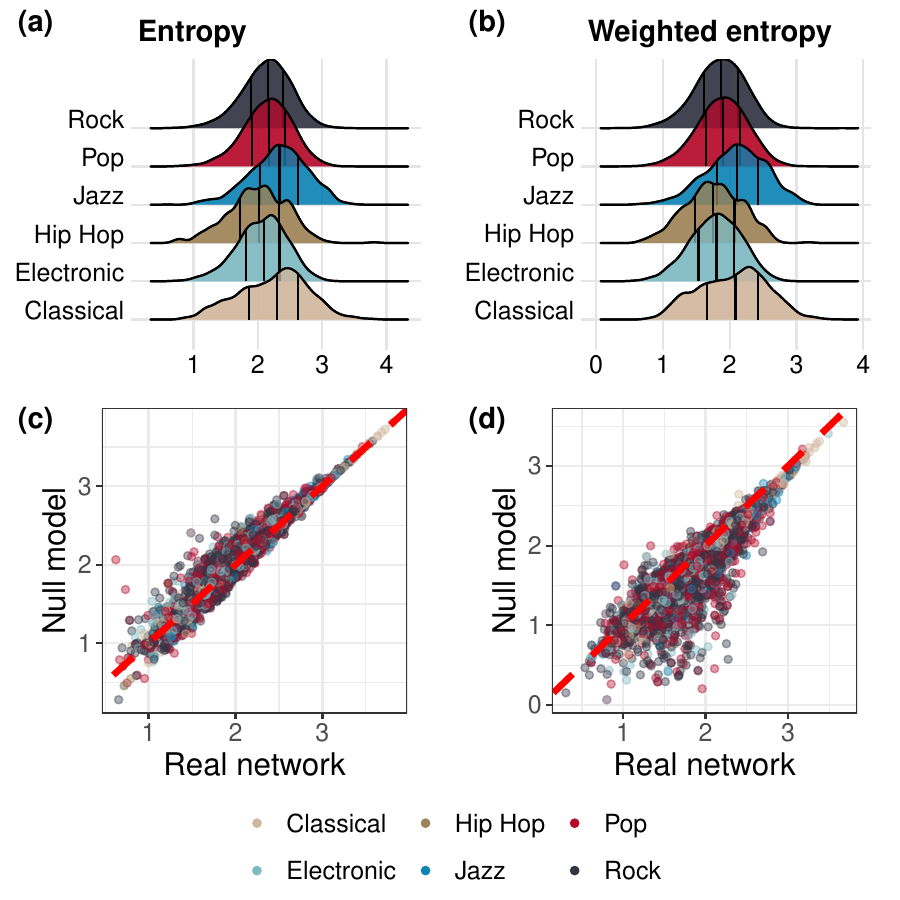}
    \caption{Distribution of $(a)$ entropy and $(b)$ its weighted counter part. 
    $(c),(d)$ shows the comparison of the values with the ones obtained from appropriately randomized versions of the networks.}
    \label{fig:entropy}
\end{figure}

Hence, we use these values to repeat the main analysis of the paper. The results are shown in Fig. \ref{fig:entropy}.

The results closely resemble the ones obtained with efficiency in the main paper, since Jazz and Classical music exhibit greater information content compared to other genres. Additionally, entropy tends to have higher values in the randomized versions of the networks, although this trend reverses when weights are taken into account.

This highlights the critical role of weights in capturing the musical properties of the network.

\begin{figure}[!ht]
    \centering
    \includegraphics[width=\linewidth]{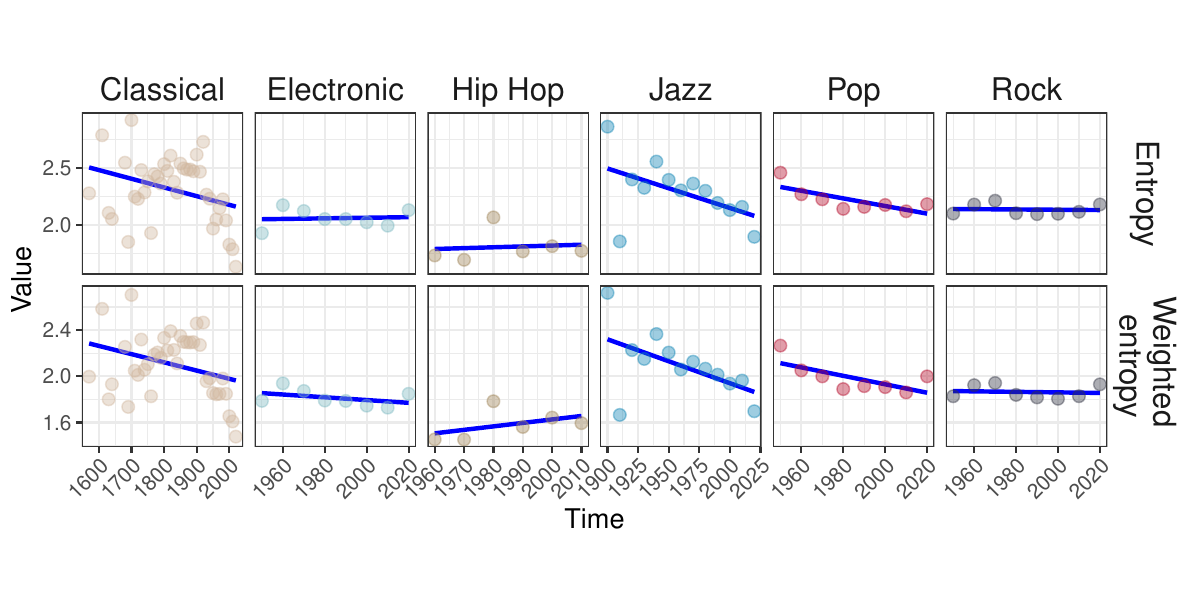}
    \caption{Trend evolution of decade-mean entropy and its weighted counterpart.}
    \label{fig:evolution_entropy}
\end{figure}

Finally, we explore the decade-mean evolution of Entropy for each genre, akin to the procedure described in the main paper. The results of the analysis are depicted in Fig. \ref{fig:evolution_entropy}.

Notably, the observed trends closely resemble the ones obtained with efficiency. In particular, Classical music shows a notable decrease in Entropy in the last century, while Jazz exhibit an initial increase followed by a decrease. The other genres maintain similar constant values. 

\subsection*{Interval embeddings}
\subsubsection*{Full UMAP dimensionality reduction}
\begin{figure}[!ht]
    \centering
    \includegraphics[width=0.7\linewidth]{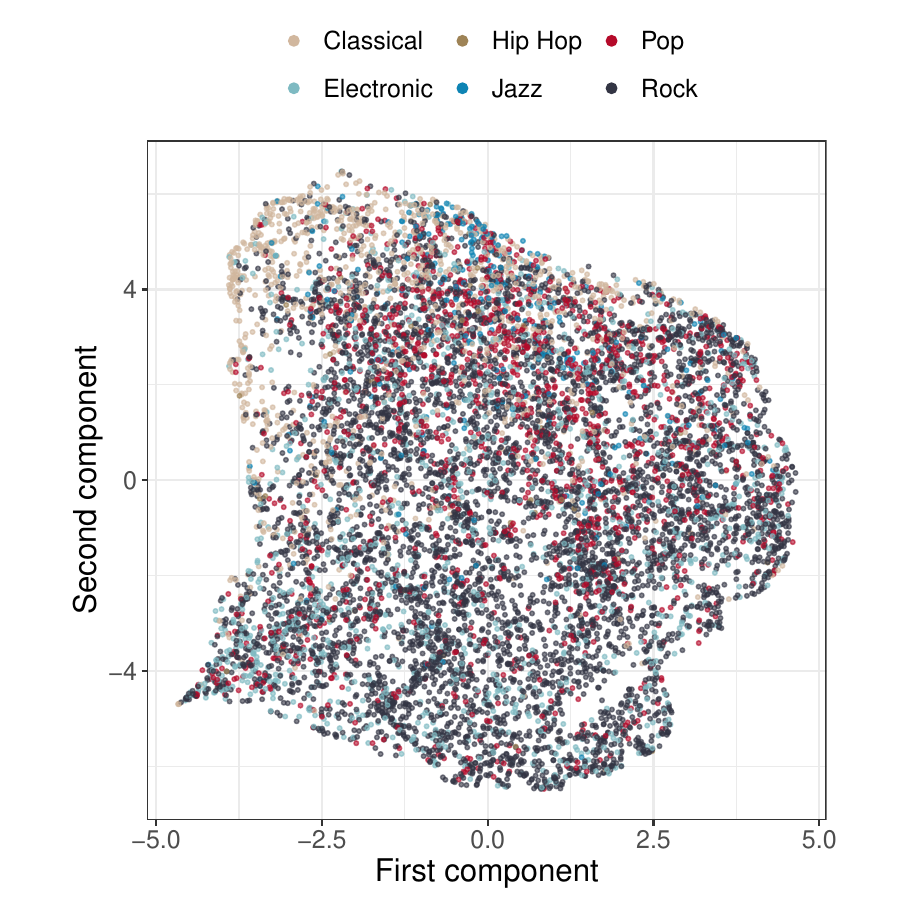}
    \caption{Low-dimensional representation of each network, with coordinates obtained from UMAP. For visualization purposes, a sample of $10^4$ points is shown.}
    \label{fig:umap_full}
\end{figure}

Figure \ref{fig:umap_full} shows the $2-$dimensional coordinates of interval embeddings obtained using UMAP.

Interestingly, a distinct cluster of Classical music emerges in the upper-left portion of the plot, while Jazz songs form a noticeable cluster in the mid-upper region. In contrast, the remaining genres exhibit more mixed behavior and lack clear clustering.

The relatively high values achieved by artists indicate limited exploration. However, this may again reflect a universal preference for specific musical patterns, as certain transitions may be generally considered dissonant.

\subsubsection{\texttt{graph2vec} embeddings}
We replicate the main analysis of the paper using embeddings generated by \texttt{graph2vec}, an algorithm that extends NLP techniques to learn networks' features and embed them into a suitable high-dimensional space.

\begin{figure}[!ht]
    \centering
    \includegraphics[width=\linewidth]{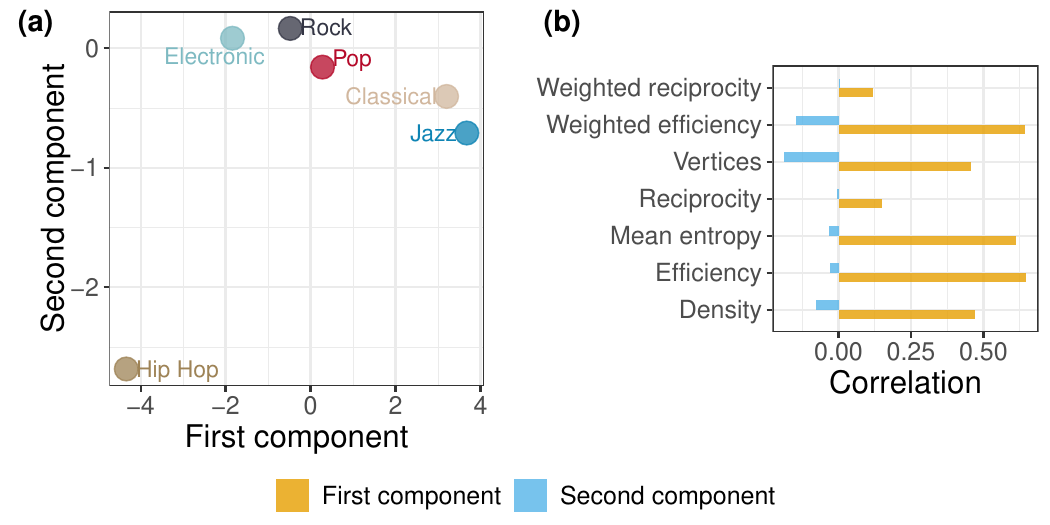}
    \caption{$(a)$ Center of mass for each genre, computed using UMAP $2-$dimensional coordinates. $(b)$ Correlation between network measures and UMAP coordinates.}
    \label{fig:graph2vec_genres}
\end{figure}

Figure \ref{fig:graph2vec_genres}$(a)$ shows the center of mass of each genre according to the $128-$dimensional embeddings created by the algorithm. 

Notably, we observe a small distance between Classical and Jazz music. Further, in this case, Hip Hop tends to cluster away from the other genres.

To interpret the role of components, panel $(b)$ shows the correlation between measures and coordinates. In this case, the first component can be interpreted as a mixture of topological and weighted properties of the network, while the second is more hardly interpretable. Notably, the high values of Classical and Jazz on the first component confirm again their higher complexity. 

Finally, we observe a negative correlation between the $GS-$score and the artists' popularity ($r = -0.12, p < 0.001$) as in the main paper. 

\begin{figure}[!ht]
    \centering
    \includegraphics[width=0.5\linewidth]{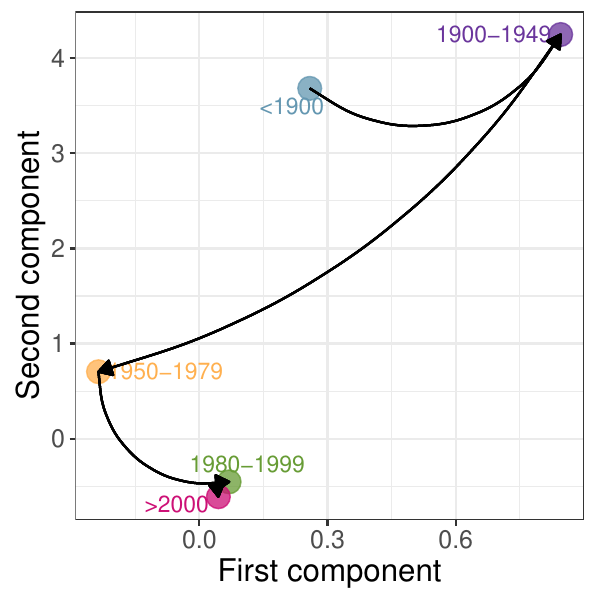}
    \caption{Center of mass of each musical period, obtained using UMAP on embeddings created starting from \texttt{graph2vec}.}
    \label{fig:umap_time_graph2vec}
\end{figure}

Shifting to the time analysis, Fig. \ref{fig:umap_time_graph2vec} depicts the center of mass of each musical period, as reported in the main paper. 

In this case as well, we observe a shift between music composed before $1950$ and more recent pieces, which show greater similarity.

\subsubsection*{Comparison with null embeddings}
To check the robustness of the interval embeddings in separating music from noise, we select a random sample of $1000$ networks having at least $30$ nodes and we randomize them by rewiring edges and randomly assigning weights. We then construct again the embeddings, thus obtaining a set ${\bf v}_{null}$ of vectors associated with randomized networks. These randomized networks serve as representations of noise, lacking any meaningful musical structure.

Then, we apply UMAP to the matrix containing these embeddings and their real counterpart.

\begin{figure}[!ht]
    \centering
    \includegraphics[width=0.5\linewidth]{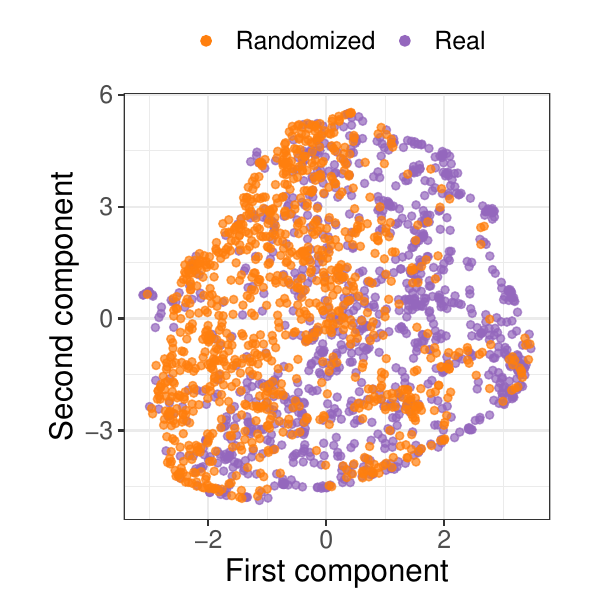}
    \caption{Results of UMAP applied a set of embeddings and their counterpart obtained after randomization of networks.}
    \label{fig:umap_null_model}
\end{figure}

Their $2-$dimensional representation is depicted in Figure \ref{fig:umap_null_model}. 

Interestingly, we can observe two distinct clusters with a lower number of intersections. This corroborates the efficacy of our embeddings in capturing relevant musical properties of the networks. 

\subsection*{Time analysis}
\subsubsection*{Robustness of release date estimation}
As detailed in the main paper, we use Gemini to estimate the release date for each song in our dataset, as Spotify's release date information is often inaccurate.

Although innovative, LLM may assign wrong release dates. However, without a ground truth, it is not easy to estimate the validity of our approach. 

To try to fill this gap, we select a random sample of $100$ MIDI files and manually annotate their release date. Then, we compare the original release date with that of Spotify and Gemini.

In particular, as in the main paper, we associate each song to one between $5$ musical periods, namely $<1900,1900-1949,1950-1979,1980-1999,>2000$ according to its release date.

\begin{figure}[!ht]
    \centering
    \includegraphics[width=0.7\linewidth]{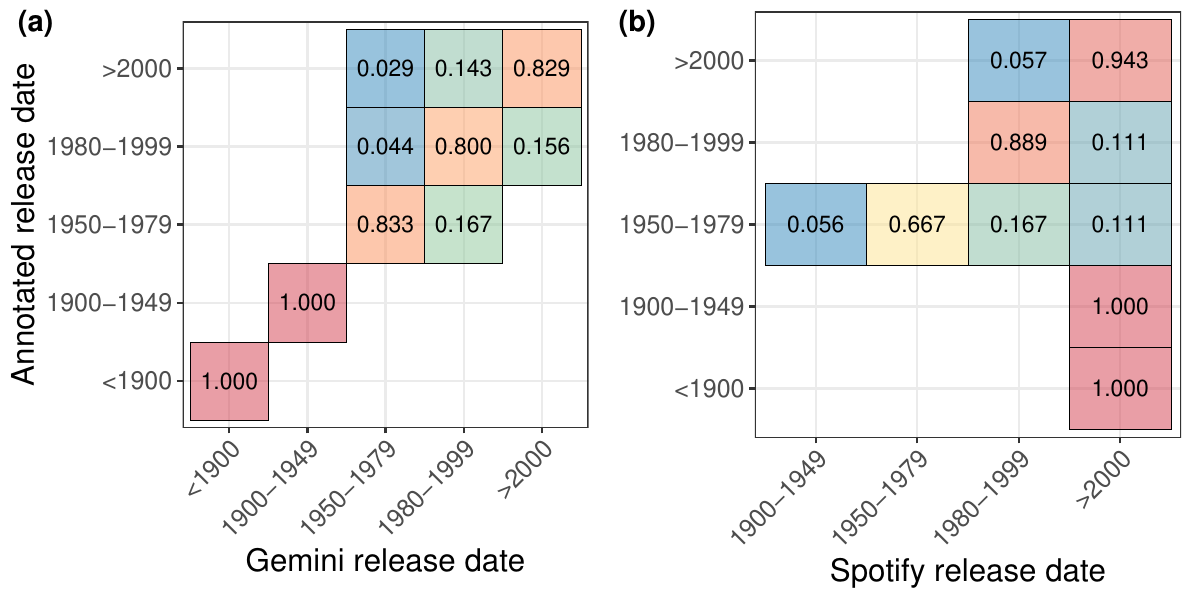}
    \caption{Contingency matrix illustrating the distribution of songs from a given musical era as classified by $(a)$ Gemini and $(b)$ Spotify.}
    \label{fig:release_date_robustness}
\end{figure}

Figure \ref{fig:release_date_robustness} presents the contingency matrices comparing the results of $(a)$ Gemini and $(b)$ Spotify in predicting the real release period of a song in our sample. 

While Spotify tends to correctly associate release dates of recent songs, i.e. from 1980 onward, older songs are better predicted by Gemini, which significantly outperforms Spotify in this regard. Notably, Spotify reports no songs released before 1900.

For these reasons, we have opted to use Spotify's release date for songs released after 1980, and Gemini's release date for those before, leveraging the strengths of both approaches.

\subsubsection*{Components interpretation}
\begin{figure}[!ht]
    \centering
    \includegraphics[width=0.8\linewidth]{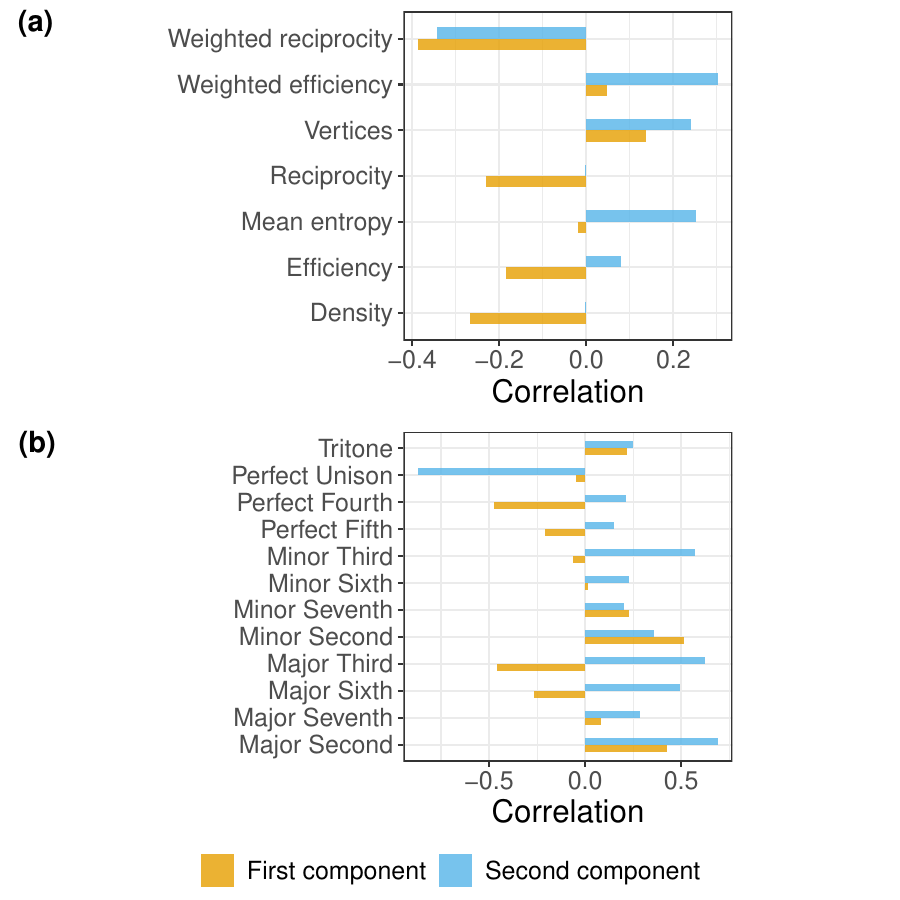}
    \caption{$(a)$ correlation between measures and components. $(b)$ correlation between intervals components and coordinates.}
    \label{fig:cor_time_embeddings}
\end{figure}

Akin to the procedure described in the main paper, Figure \ref{fig:cor_time_embeddings} shows the correlation between the two components obtained using UMAP and the $(a)$ network measure or $(b)$ intervals components. 

\subsubsection*{Result of Mann-Kendall}
In this section we report the results of the Mann-Kendall test applied to the time-trends depicted in the main paper. The results are summarized in Table \ref{tab:mk_trend}.

\begin{table}[!ht]
\centering
\begin{tabular}{llrr}
  \hline
Genre & Measure & $\tau$ & $p_{adj}$ \\ 
  \hline
Classical & Efficiency & -0.401 & 0.001 \\ 
  Classical & Weighted 
efficiency & -0.263 & 0.019 \\ 
  Electronic & Efficiency & -0.143 & 0.711 \\ 
  Electronic & Weighted 
efficiency & -0.429 & 0.347 \\ 
  Hip Hop & Efficiency & 0.600 & 0.266 \\ 
  Hip Hop & Weighted 
efficiency & -0.067 & 1.000 \\ 
  Jazz & Efficiency & -0.487 & 0.048 \\ 
  Jazz & Weighted 
efficiency & -0.256 & 0.246 \\ 
  Pop & Efficiency & 0.143 & 0.711 \\ 
  Pop & Weighted 
efficiency & -0.500 & 0.216 \\ 
  Rock & Efficiency & 0.714 & 0.037 \\ 
  Rock & Weighted 
efficiency & -0.357 & 0.266 \\ 
   \hline
\end{tabular}
\caption{Result of Mann-Kendall tests applied to the trends of efficiency evolutions. The $p-$values have been corrected using the Holm procedure.}
\label{tab:mk_trend}
\end{table}

\subsection*{Analysis with popularity}
\subsubsection*{Measures and popularity}
In the main paper, we examined the characteristics of networks derived from different musical genres. However, we have yet to consider whether these measures are related to the popularity of the songs, and if so, how. 

To explore this dimension, we compute the correlations between measures and tracks' popularity, gathered using Spotify API. 
The results of the analysis are depicted in Figure \ref{tab:correlations}, which reports the Pearson correlations between measures and popularity, divided according to the musical genres. Correlations with all the measures are reported in SI.

\begin{table}[ht]
\centering
\begin{tabular}{llrr}
  \hline
Genre & Measure & $p_{adj}$ & $r$ \\ 
  \hline
Classical & Density & $<0.001$ & -0.121 \\ 
 Classical & Mean entropy & $<0.001$ & -0.166 \\ 
  Classical & Weighted reciprocity & $<0.001$ & 0.088 \\ 
  Classical & Weighted efficiency & $<0.001$ & -0.192 \\ 
  Electronic & Density & 0.004 & 0.052 \\ 
  Electronic & Mean entropy & 0.356 & -0.021 \\ 
  Electronic & Weighted reciprocity & 0.097 & 0.034 \\ 
  Electronic & Weighted efficiency & 0.356 & -0.020 \\ 
  Hip Hop & Density & 1.000 & 0.019 \\ 
  Hip Hop & Mean entropy & 1.000 & -0.004 \\ 
  Hip Hop & Weighted reciprocity & 0.108 & -0.106 \\ 
  Hip Hop & Weighted efficiency & 1.000 & -0.014 \\ 
  Jazz & Density & 0.007 & -0.068 \\ 
  Jazz & Mean entropy & $<0.001$ & -0.161 \\ 
  Jazz & Weighted reciprocity & $<0.001$ & 0.158 \\ 
  Jazz & Weighted efficiency & $<0.001$ & -0.131 \\ 
  Pop & Density & $<0.001$ & 0.044 \\ 
  Pop & Mean entropy & $<0.001$ & -0.040 \\ 
  Pop & Weighted reciprocity & $<0.001$ & 0.049 \\ 
  Pop & Weighted efficiency & $<0.001$ & -0.036 \\ 
   Rock & Density & $<0.001$ & 0.036 \\ 
   Rock & Mean entropy & $<0.001$ & -0.042 \\ 
   Rock & Weighted reciprocity & $<0.001$ & 0.054 \\ 
   Rock & Weighted efficiency & $<0.001$ & -0.062 \\ 
   \hline
\end{tabular}
\caption{Pearson correlations between networks' measures and Spotify track popularity. The $p-$values were adjusted using the Holm correction.}
\label{tab:correlations}
\end{table}

Although most cases reveal only weak correlations, we can still uncover some interesting insights. It is worth noting that all values are significant, except for those associated with the Hip Hop and Electronic genres.

The results highlight that, in Jazz and Classical music, more popular songs are the ones having lower density, i.e. lower number of transitions between notes.
We find a similar pattern for mean entropy, where all genres exhibit negative correlations.

For what regards reciprocity and efficiency, we observe exactly opposite patterns. In fact, the former shows positive correlations with popularity, while the latter has negative correlations. 

Observed together, the results point to an inverse relationship between popularity and musical complexity.
Notably, this holds especially in Jazz and Classical music, which have been proven to show greater complexity compared to the other genres.

\subsubsection*{Artist Exploration and Diversity in Musical Composition}
The embeddings also allow us to examine how artists use musical intervals across their discography. To achieve this, we employ a metric called the Generalist-Specialist (GS) score.

The $GS$-score measures the concentration of a set of points in a high dimensional space, and it has been recently employed in multiple works \cite{anderson2020algorithmic,mok2022dynamics}.
Let us denote with $S = \{s_1, \ldots, s_n\}$ a set of vectors in a vectorial space $V$. The $GS-$score is defined as

\begin{equation}
    GS = \frac{1}{n} \sum_{j = 1}^n \frac{s_j \cdot \mu}{||s_j|| \cdot || \mu||}
\end{equation}

where $\mu$ is the center of mass of $S$. Note that $GS$ is the average of the cosine similarity between each $s_j$ and their center of mass $\mu$. Therefore, it measures the expected similarity between the vectors and their center of mass. Values close to $1$ indicate a high concentration of the vectors, which lie close to their center of mass. On the other hand, values close to $0$ indicate vectors spread apart in the space.

In our context, this measure quantifies an artist's extent of exploration of different music intervals by measuring the concentration of their tracks' embeddings within the $12$-dimensional interval space (see Methods for further details). 
Recall that a high score indicates an artist with a low rate of exploration, i.e. who creates similar music, while lower values indicate greater diversity. 

\begin{figure}[!ht]
    \centering 
    \includegraphics[width=0.8\linewidth]{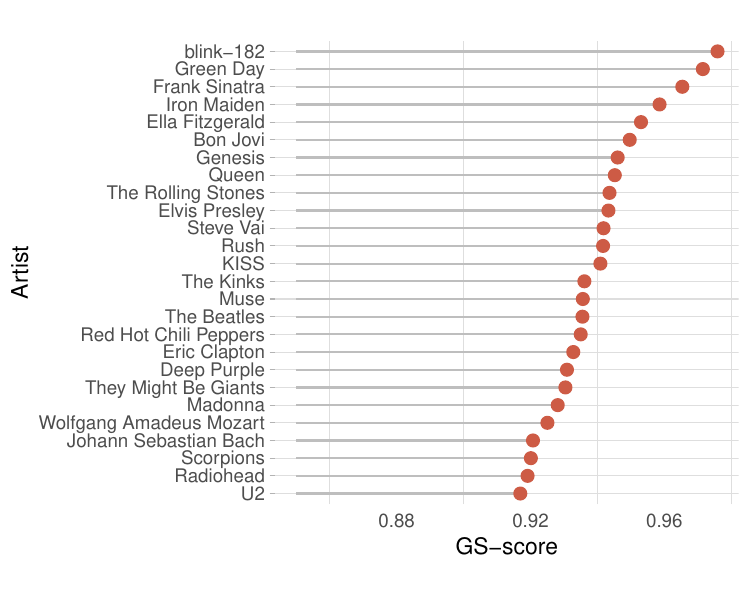}
    \caption{$GS-$score of artists with at least 80 songs in our dataset.}
    \label{fig:gs_score}
\end{figure}

Generally, all artists obtain high values, probably due to the existence of musical patterns forcing a limited exploration of the space.
To give an idea of the values obtained by artists, Fig. \ref{fig:gs_score} shows the $GS-$score obtained for artists having at least $80$ songs in our dataset. 

The same index allows for studying the relationship between artists' popularity and musical diversity. To accomplish this, we calculate the correlations between the $GS$-score and Spotify popularity for all artists with at least $5$ MIDI tracks in our dataset. The analysis reveals a slight but statistically significant negative correlation between the two variables ($r = -0.12, p < 0.001$), suggesting that the more popular artists tend to exhibit greater musical diversity.

\bibliography{bibliography}

\end{document}